\documentclass[a4paper]{article}

\usepackage{INTERSPEECH2020}
\newcommand{\argmax}{\operatornamewithlimits{argmax}}
\usepackage{caption}
\usepackage{subcaption}
\usepackage{amssymb}
\usepackage{multirow}
\usepackage{float}
\usepackage{lipsum}
\usepackage{hyperref}
\usepackage[capitalize]{cleveref}
\Crefname{chapter}{Chap.}{Chaps.}
\Crefname{section}{Sec.}{Secs.}
\Crefname{figure}{Fig.}{Figs.}

\newcommand\uno{1}
\newcommand\due{2}
\newcommand\tre{3}
\newcommand\quattro{4}
\newcommand\cinque{5}
\newcommand\sei{6}

\usepackage{pifont}
\newcommand{\cmark}{\ding{51}}%

\usepackage{array}
\newcolumntype{?}{!{\vrule width 1.5pt}}

\title{Context-Dependent Acoustic Modeling without Explicit Phone Clustering}
\name{Tina Raissi, Eugen Beck, Ralf Schl\"uter, Hermann Ney}
\address{
	Human Language Technology and Pattern Recognition Group\\
	RWTH Aachen University}
\email{\{raissi,beck,schlueter,ney\}@cs.rwth-aachen.de}

\begin{document}
\maketitle
\begin{abstract}
	
Phoneme-based acoustic modeling of large vocabulary automatic speech recognition takes advantage of phoneme context.\ The large number of context-dependent~(CD) phonemes and their highly varying statistics require tying or smoothing to enable robust training.\ Usually, classification and regression trees are used for phonetic clustering, which is standard in hidden Markov model~(HMM)-based systems.\ However, this solution introduces a secondary training objective and does not allow for end-to-end training.\ In this work, we address a direct phonetic context modeling for the hybrid deep neural network~(DNN)/HMM, that does not build on any phone clustering algorithm for the determination of the HMM state inventory.\ By performing different decompositions of the joint probability of the center phoneme state and its left and right contexts, we obtain a factorized network consisting of different components, trained jointly.\ Moreover, the representation of the phonetic context for the network relies on phoneme embeddings.\ The recognition accuracy of our proposed models on the Switchboard task is comparable and outperforms slightly the hybrid model using the standard state-tying decision trees.


\end{abstract}
\noindent\textbf{Index Terms}: automatic speech recognition, context-dependent acoustic modeling, hybrid DNN/HMM system

\section{Introduction}
\label{sec:intro}

The realization of the phonetic co-articulation effect in large vocabulary continuous speech recognition~(LVCSR) systems with standard Gaussian mixture model/hidden Markov model~(GMM/HMM) takes into account a context-dependent~(CD) representation of phones, usually triphones~\cite{bahl1989large}.\ The extension of each phoneme with its left and right contexts leads to a considerable growth of the number of possible states.\ Finding the right trade-off between the model complexity and the available data can become complicated, on the grounds that during training many triphones are unevenly distributed or never observed.\ In order to overcome sparsity issues, for long, classification and regression trees~(CART) marked the state-of-the-art in ASR for tying CD phone states into generalized triphone states~\cite{young1994tree}.\ The successful advent of neural-based models in LVCSR paved the way for the hybrid deep neural network~(DNN)/HMM architecture~\cite{bourlard2012connectionist}, where the Gaussian mixture based emission probabilities are replaced by normalized scaled generalized triphone state posteriors, predicted by a discriminative model.\ 

The introduction of CART labels as output targets of the NN model has given an important contribution to the improvement of the performance of the recognition systems, maintaining at the same time a two-fold dependency to the GMM system.\ The frame-level state alignment for training CART derives normally from a GMM.\ Furthermore, there is a mismatch between the features used for the estimation of Gaussian mixture parameters and the one used for learning the posterior probabilities of the tied-states in the neural network component.\  

The majority of the research works on the CD acoustic modeling in connection with the hybrid approach aims to either integrate the context directly into the neural network~\cite{bourlard1992cdnn,dahl2011context}, or to eliminate the dependency to the GMM system.\ It is shown that the initial alignments to the context-independent (CI) states for the standard tree-based clustering approach can be provided by a flat-started DNN~\cite{senior2014gmm}.\ Similar approaches design the set of CD targets by clustering the activations of a CI DNN~\cite{zhang2014standalone,bacchiani2014context}.\ There are also different possible training criteria for the state-tying algorithm, such as Kullback-Leibler divergence~\cite{razavi2014modeling,gosztolya2015building}, entropy~\cite{zhu2015gaussian}, and based on DNN and classification error~\cite{wiesler2010discriminative}.\ The elimination of the state-tying decision trees is the topic of research also for end-to-end models such as CTC  where a CD embedding network is applied instead~\cite{chorowski2019towards}.

The common trait between most of the mentioned works is a phone clustering principle and the necessity of having one more training and optimization step.\ It is important to underline that in addition to the supplementary time and resource effort, another crucial concern regarding this further modeling approach is how the set of clustered states can affect the decision boundaries in the final neural network, which learns the probability distribution over their posteriors.\ This is especially true when the classic phonetic decision trees are involved. The relative heuristics regarding the choice of the questions or maximum number of leaves can affect directly the definition of the set of class labels, which, if not well-defined, can lead to overfitting problems~\cite{bell2016multitask}.\

In this work, we propose a CD acoustic modeling for the hybrid approach, which disposes of the necessity of an additional phone clustering step for the determination of the HMM state inventory.\ The resulting model is partitioned into separate components, trained conjointly, and corresponding to one of the factorized elements of the joint probability of the center phoneme state with its left and right phonetic contexts.\ Depending on the type of decomposition, each component learns a posterior probability distribution over phonemes and phoneme states in mono-, di- and triphone context.\ We show that for the Switchboard task the recognition system built upon our direct context integration approach with no state-tying clustering can obtain a similar performance to a hybrid model using standard tying based on CART.\

\section{Formulation of the Problem}
\label{sec:prob}

The statistical formulation of automatic speech recognition task maximizes the a-posteriori probability of a word sequence $w_1^N$ of length $N$ given the acoustic feature sequence $x_1^T$ of length $T$, with $T \gg N$, based on Bayes decision rule~\cite{bayes1763lii}:
\vspace{-0.1cm}
\begin{equation}
\label{eq:bayes}
x_1^T \rightarrow\tilde{w}_1^N(x_1^T) = \underset{w_1^N}{\argmax} \left\lbrace p(x_1^T|w_1^N) \cdot p(w_1^N) \right\rbrace \vspace{-0.1cm}
\end{equation}
The acoustic-phonetic component $p(x_1^T|w_1^N)$ of \cref{eq:bayes}, in the standard HMM with generative approach and involving a sequence of triphone states $s_1^T$ is formulated as:
\vspace{-0.1cm}
\begin{eqnarray*}
	p(x_1^T|w_1^N)\hspace{-0.2cm}&=& \hspace{-0.2cm}\sum_{s_1^T}\prod_{t=1}^{T} p( x_t|s_t, w_1^N)\cdot p(s_t|s_{t-1}, w_1^N)\\
	\hspace{-0.2cm}&=& \hspace{-0.2cm}\sum_{s_1^T}\prod_{t=1}^{T} p( x_t|s_t, \phi_1^M, w_1^N )\cdot p(s_t|s_{t-1}, \phi_1^M, w_1^N ) 
\end{eqnarray*}
where $\phi_1^M$ represents a suitable triphone sequence of length $M$ corresponding to the word sequence.

\section{Integration of the Context}

Denote by $\left\lbrace \phi_{\ell},\phi_{c},\phi_{r}\right\rbrace_t$ the set of left, center and right phonemes of the aligned triphone at time frame $t$.\ Each phoneme consists of three HMM states, and each state can be associated with a state class $c(s_t, w_1^N) = \left\lbrace \phi_{\ell},\phi_{c},\phi_{r}, i \right\rbrace _t$, where $i$ enumerates the HMM state of the corresponding triphone.\ The likelihood of generating a feature vector $x$ at time frame $t$ given a triphone, can be written as:
\begin{equation}
\nonumber
p(x_t|s_t, \phi_1^M, w_1^N)=p(x_t | c(s_t, w_1^N)) =p_{_t}(x| \phi_{\ell},\phi_{c},\phi_{r},i) 
\end{equation}
For simplicity, we use the parametrized probability distribution $p$ and its further denotation $p_{_t}$ at time frame $t$, interchangeably.\ By applying Bayes identity we have:
\begin{equation}
\label{eq:bayesid}
p(x | \phi_{\ell},\phi_{c},\phi_{r}, i)=\frac{p(\phi_{\ell},\phi_{c},\phi_{r}, i | x)\cdot p(x)}{p(\phi_{\ell},\phi_{c},\phi_{r},i)}  
\end{equation}
Let $\sigma_c$ be the current HMM state within the center phoneme, the CD neural network should ideally model the joint probability of $\sigma_c$ with the left and right phonetic contexts appearing in the nominator of \cref{eq:bayesid}, which can be written through the following mapping as:
\begin{equation}
\label{eq:sigma_c}
p(\phi_{\ell},\phi_{c},\phi_{r}, i | x) \rightarrow p(\phi_{\ell},\sigma_c,\phi_{r} | x)  
\end{equation}

\section{Different Decompositions} \label{sec:deomp}

The joint posterior probability distribution of \cref{eq:sigma_c} would demand a high number of parameters and an infeasible memory requirement, if conceived as the output of a neural network.\ One possible solution is to obtain a factorization into CD probabilities, by applying the classic Markov chain rule~\cite{morgan1992factoring}.\ 


\subsection{Diphone}

The emission probability defined for the diphone model, as shown in \cref{fig:diphone}, is obtained by conditioning only on the left phonetic context.\ Starting with the modified version of \cref{eq:bayesid}, which takes also into consideration the mapping  \cref{eq:sigma_c} we have:
\begin{subequations}
	\label{eq:dph}
	\begin{align}
	p(x | \sigma_c,\phi_{\ell})&=\frac{p(\sigma_c, \phi_{\ell}| x)\cdot p(x)}{p(\sigma_c,\phi_{\ell})} \nonumber \\
	&=\frac{p(\sigma_c|\phi_{\ell}, x)\cdot p(\phi_{\ell}|x)\cdot p(x)}{p(\sigma_c|\phi_{\ell})\cdot p(\phi_{\ell})} \tag{4}
	\end{align}
\end{subequations}

\subsection{Triphone Forward}

\begin{figure}[t]
	\centering
\subcaptionbox{Diphone. \label{fig:diphone}}{%
	\includegraphics[width=0.36\columnwidth, trim={19cm 0cm 0cm 13cm},clip]{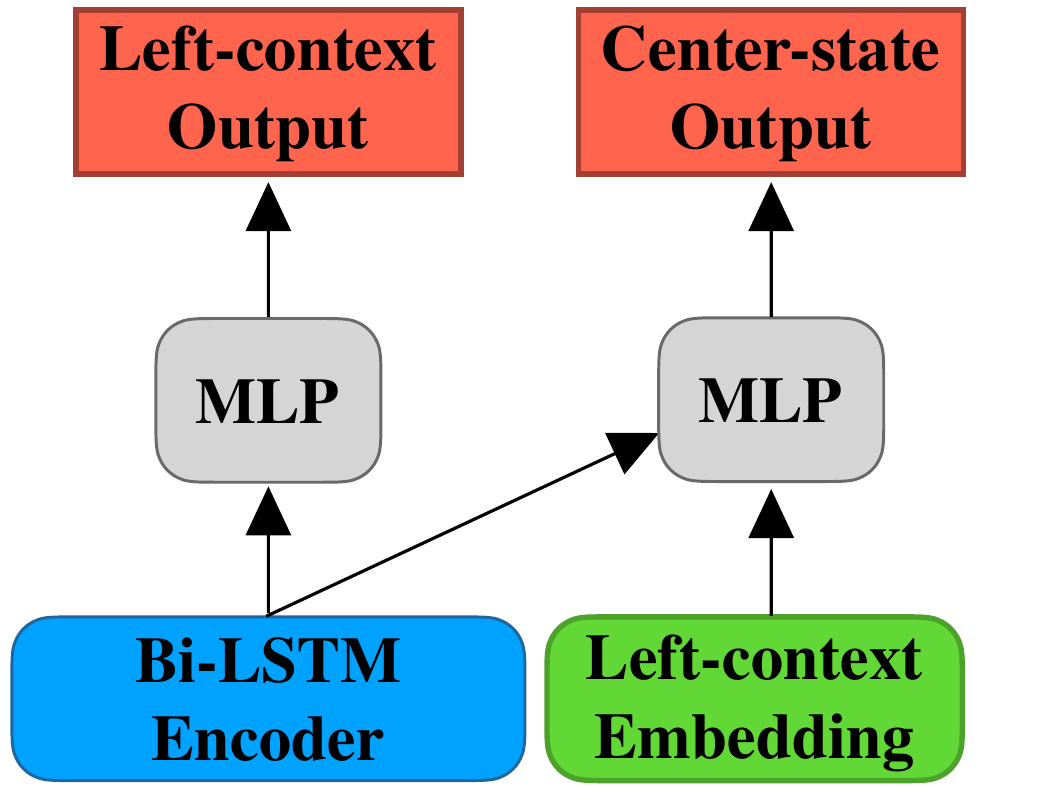}%
}\qquad
\subcaptionbox{Triphone forward.\label{fig:tri-fwd}}{%
		\includegraphics[width=0.53\columnwidth, trim={0cm 0cm 14cm 13cm},clip]{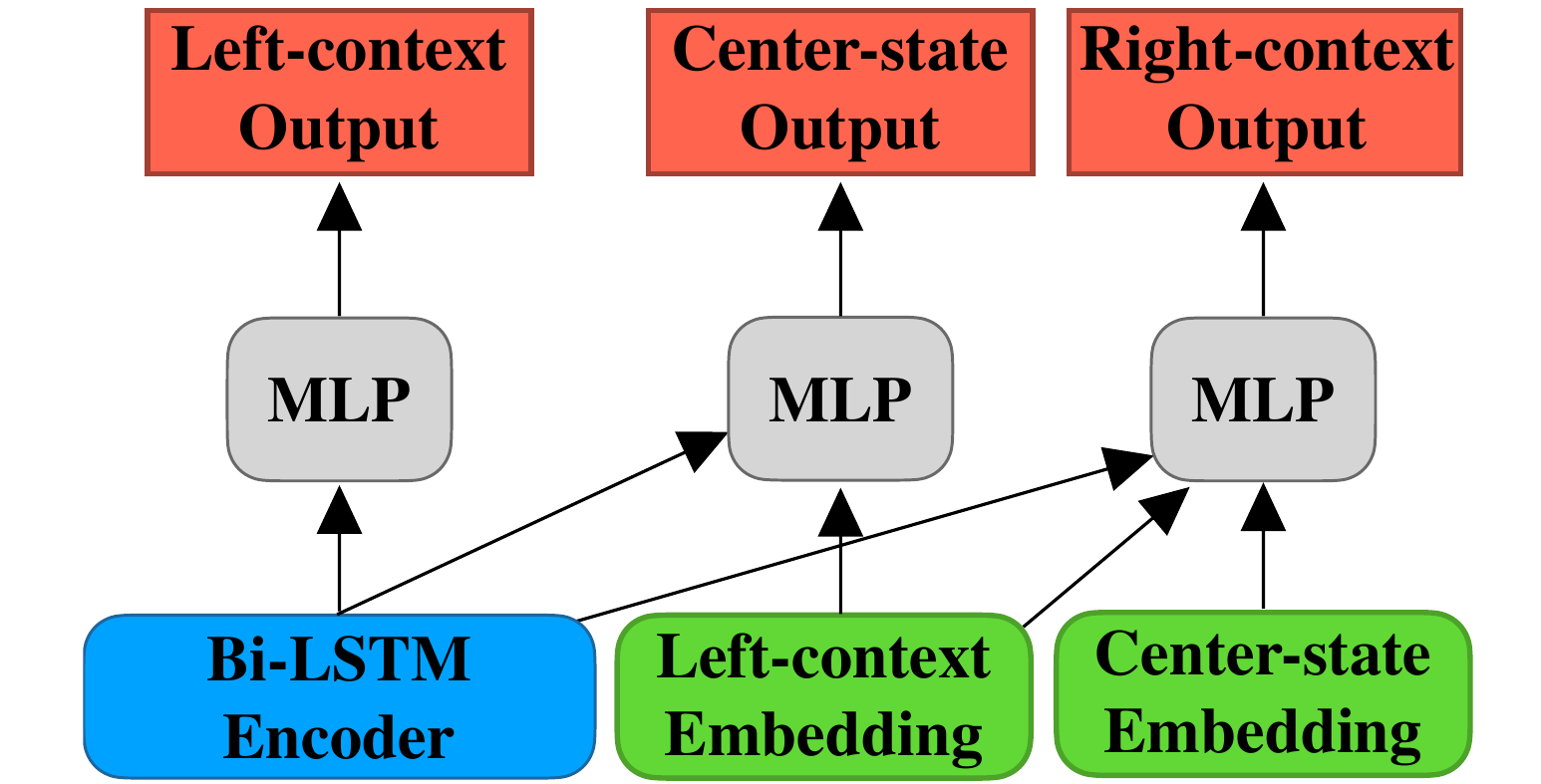}
}
\subcaptionbox{Triphone symmetric \\ (simplified).	\label{fig:tri-sym}}{%
	\hspace{-0.35cm}		
	\includegraphics[width=0.55\columnwidth, trim={13cm 0cm 0cm 13.1cm},clip]{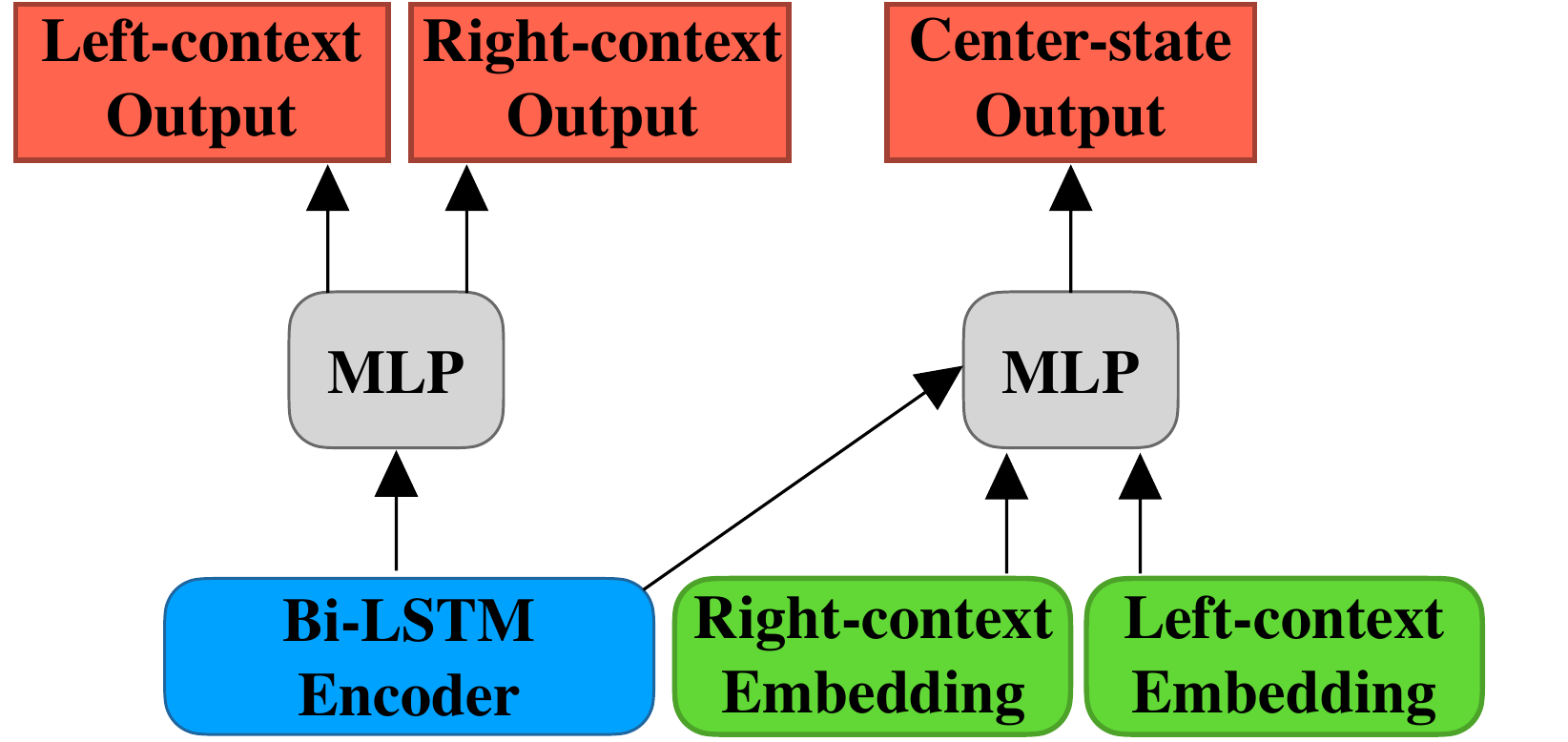}%
}\hspace{-0.57cm}
\subcaptionbox{Triphone backward.\label{fig:tri-bwd}}{%
	\includegraphics[width=0.535\columnwidth, trim={0cm 0cm 13.5cm 12cm},clip]{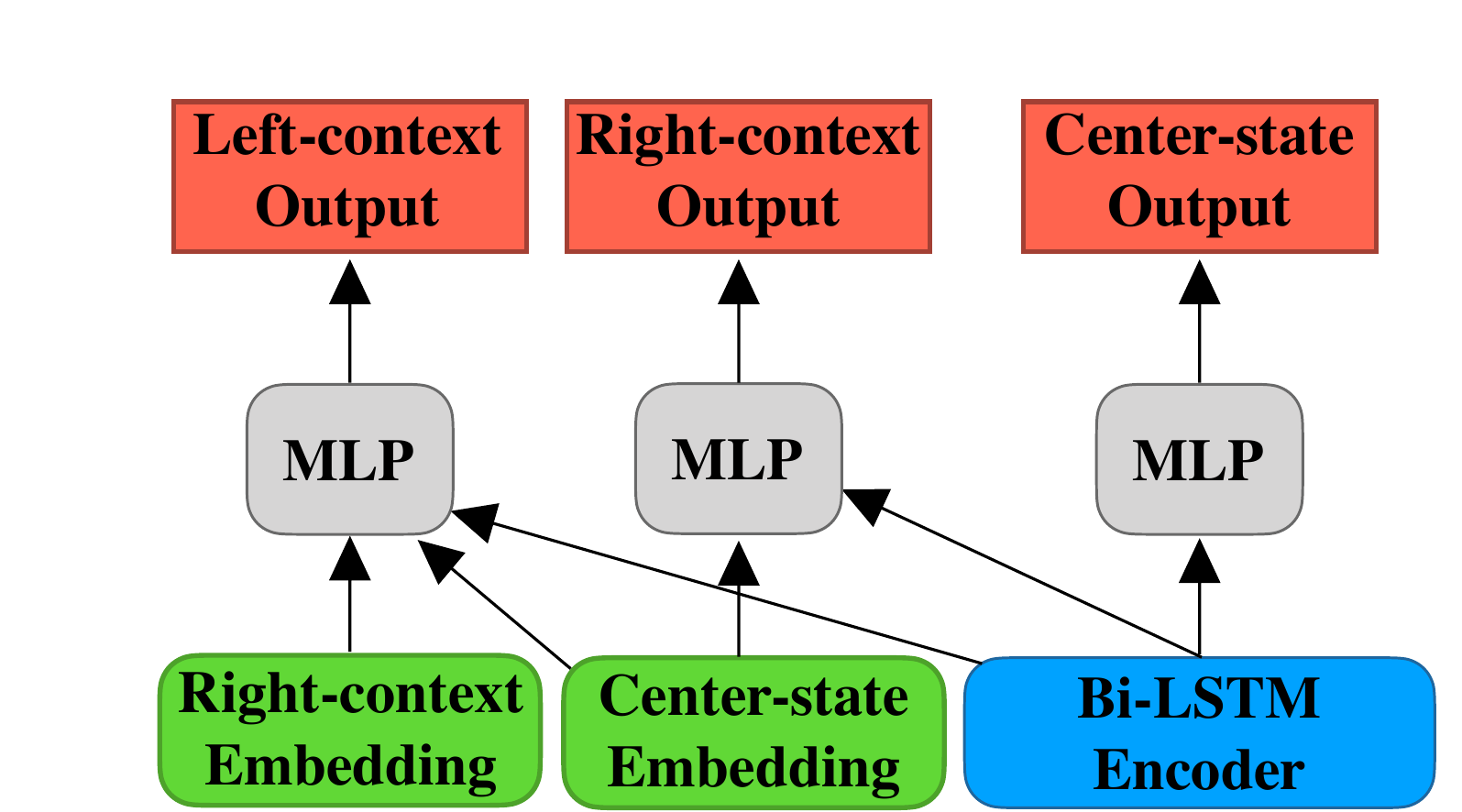}
}
	
	\caption{The architecture of different models defined in \cref{eq:dph,eq:tri-fwd,eq:trisymm,eq:tri-bwd}.\ The left and right output layers have the respective phoneme identities $\phi_\ell$ and $\phi_r$ as targets.\ The target for the center phoneme output is the CI state referred to the phoneme inventory of the vocabulary.}
	\label{fig:archs}
	\vspace{-0.5cm}	
\end{figure}

In case of all triphone models, depicted in \cref{fig:tri-fwd,fig:tri-sym,fig:tri-bwd}, it is possible to achieve different decompositions by having as the start point all three entities, namely right and left contexts along with the center phoneme state.\
beginning with the right context, the chain rule will produce a left-to-right trigram, as below:
\begin{subequations}
	 \label{eq:tri-fwd}
	 \begin{align}
	p(x | \phi_{\ell},\sigma_c,\phi_{r}) &= \frac{p(\phi_{\ell},\sigma_c, \phi_{r} |x)\cdot p(x)}{p(\phi_{\ell}, \sigma_c,  \phi_{r})} \nonumber \\
	&= \frac{p(\phi_{r}|\phi_{\ell},\sigma_c, x)\cdot p(\sigma_c|\phi_{\ell}, x)\cdot p(\phi_{\ell}|x) \cdot p(x)}{p(\phi_{r} | \phi_{\ell},\sigma_c)\cdot  p(\sigma_c|\phi_{\ell})\cdot p(\phi_{\ell})}\tag{5}
	\end{align}	
\end{subequations}

\begin{table*}[t]	
	\setlength{\tabcolsep}{0.7em}\renewcommand{\arraystretch}{1.2}  
	\centering
	\caption{Different pre-training procedures for the proposed triphone model with forward decomposition~(Fwd) of \cref{fig:tri-fwd}.\ The outputs of the trained architectures at monophone and diphone stages, depicted in \cref{fig:mono-fwd,fig:di-fwd}, are marked in the respective network columns.\ Decoding for diphone models and Fwd follows \cref{eq:dph,eq:tri-fwd}, respectively.\ The recognition results in terms of word error rate~(WER) are over 300h Switchboard using 4-gram language model.\ The experiments of each row can be described as follows: ($\uno$) - Baseline triphone model with  standard state-tying, ($\due$) - Fwd with no pre-training, ($\tre$) - Diphone model of \cref{fig:diphone} trained with no pre-training and used for the initialization of Fwd, ($\quattro$) - Pre-trained Fwd with only monophone stage, ($\cinque$) Fwd with monophone and diphone pre-training stages, ($\sei$): Similar to experiment ($\cinque$) with optional inclusion of the network branch having output distribution\ $p(\phi_r|\sigma_c, x)$.\ }
	\label{tab:pretrain}
		\vspace{-2mm}		
	
	\begin{tabular}{|c|c?c|c|c|c|c|c|c?c|} 
		\hline			
		\multirow{3}{*}{\textbf{\#}} & \multirow{3}{*}{\textbf{Model}} &
		\multicolumn{3}{c|}{\textbf{Monophone stage}}& 
		\multicolumn{4}{c?}{\textbf{Diphone stage}} &
		\textbf{Triphone stage} \\ \cline{3-10}  
		& & \multicolumn{3}{c|}{\textbf{Network }} & \multicolumn{3}{c|}{\textbf{Network }} &\multirow{2}{*}{\textbf{[\%WER]}}&\multirow{2}{*}{\textbf{[\%WER]}}\\ \cline{3-8} 
		& & {\hskip 0.15in $\boldsymbol{\phi_{\ell}}$ \hskip 0.15in}   &   {\hskip 0.15in $\boldsymbol{\phi_{c}}$ \hskip 0.15in} & {\hskip 0.15in $\boldsymbol{\phi_{r}}$ \hskip 0.15in} & {\hskip 0.15in $\boldsymbol{\phi_{\ell}}$ \hskip 0.15in} & {\hskip 0.15in $\boldsymbol{\phi_{c}}$ \hskip 0.15in} & {\hskip 0.15in $\boldsymbol{\phi_{r}}$ \hskip 0.15in} &  & \\ 	\hline \hline
		$\uno$ & Base &\multicolumn{7}{c?}{not applicable~(n/a)}&  13.9 \\  \hline
	$\due$ & \multirow{5}{*}{Fwd} & -& -& -  & -&-&- & n/a &   13.9 \\ 	 \cline{1-1} \cline{3-10} 
		$\tre$ &  & -& -& - & \cmark & \cmark &- & 14.8 & 13.9\\ 	\cline{1-1} \cline{3-10} 
		$\quattro$ &  & \cmark& \cmark& \cmark & - & - &- & n/a &  13.9 \\ 	\cline{1-1} \cline{3-10} 
		$\cinque$ &  & \cmark& \cmark& \cmark & \cmark & \cmark &- & 14.4 & 13.8 \\ 	\cline{1-1} \cline{3-10} 
		
		$\sei$ &  & \cmark& \cmark& \cmark & \cmark&\cmark&\cmark & 14.2 &  13.6 \\ 	\hline
		
	\end{tabular}   
	
\end{table*} 

\begin{figure*}[t]	
		\vspace{-0.3cm}	
			\begin{subfigure}{0.05\linewidth}
				\hspace{0.5cm}

			\end{subfigure}
	\begin{subfigure}{0.45\linewidth}
		\centering
		\includegraphics[width=0.65\linewidth, trim={8cm 6cm 7cm 6cm},clip]{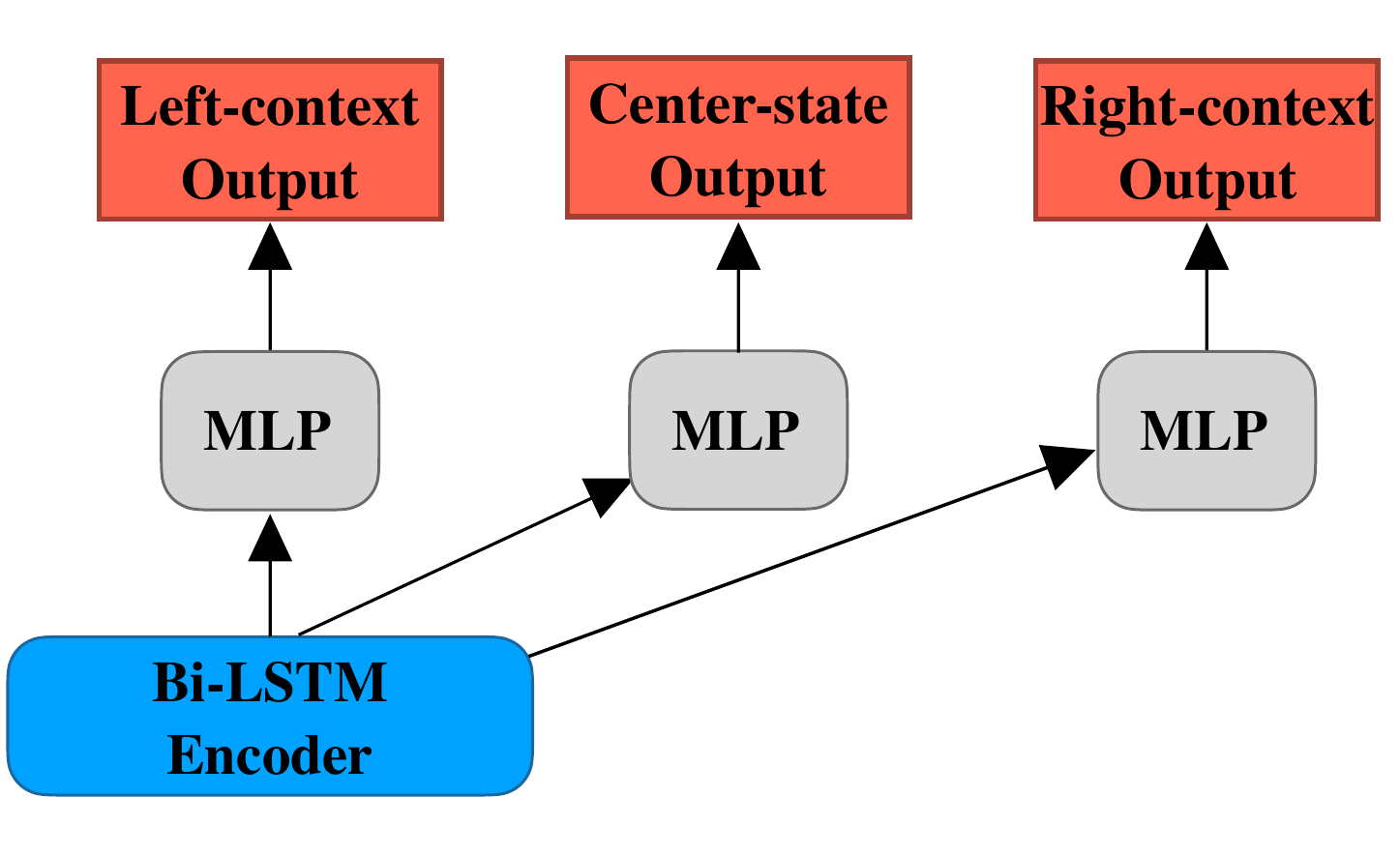}
		\caption{Monophone stage of rows $\quattro$ to $\sei$.}
		\label{fig:mono-fwd}
	\end{subfigure}
	\hspace{-2.7cm}
	\begin{subfigure}{0.45\linewidth}
		\centering
		\includegraphics[width=0.7\linewidth, trim={8cm 6cm 6cm 6cm},clip]{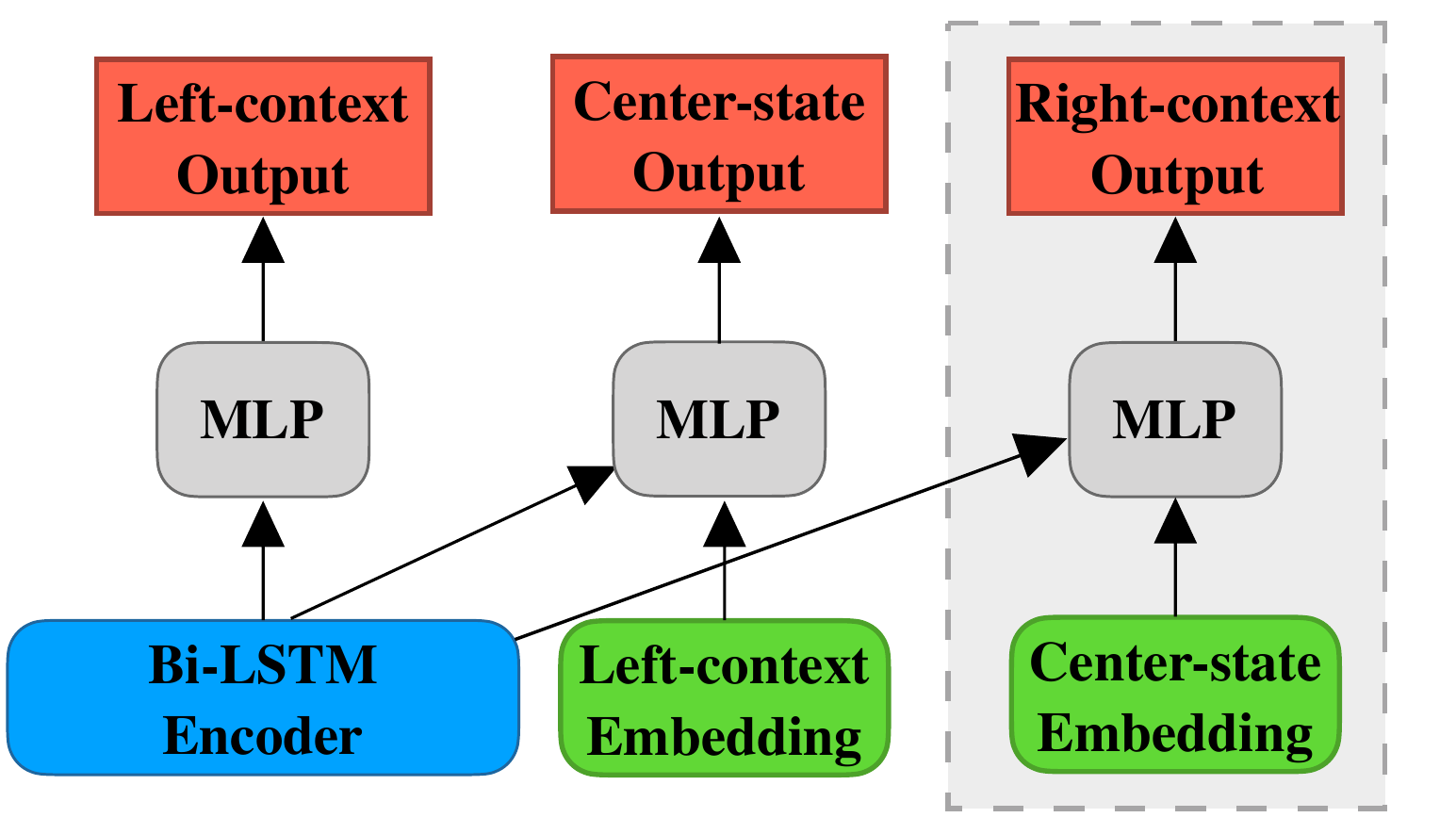}
		\caption{Diphone stage rows $\tre$, $\cinque$ and $\sei$.}
		\label{fig:di-fwd}
	\end{subfigure}

		\vspace{-2mm}

	\caption{The network architectures used in monophone and diphone stages for pre-training of proposed triphone model with forward decomposition.\ The highlighted branch of the diphone network in \cref{fig:di-fwd} with output distribution $p(\phi_r|\sigma_c, x)$ is included only for Experiment $\sei$ of \cref{tab:pretrain}.}
	\label{fig:fwd-archs}
	
	\vspace{-4mm}
	
\end{figure*}

		\vspace{-0.7cm}
\subsection{Triphone Symmetric}

Another possible decomposition starts with the center phoneme state given the left and right contexts.\ The context-dependency in the remaining factors is not taking into account the center phoneme.\ By assuming that there is no interdependency between the left and right contexts, we drop the dependency to the right context $\phi_{r}$ in the probability value $p(\phi_{\ell} | \phi_r, x)$ of \cref{eq:trisymm-a}, ending up with \cref{eq:trisymm-b}.\ This independence assumption is valid also for the respective prior $p(\phi_{\ell} | \phi_r)$.\ 
\begin{subequations}
		\label{eq:trisymm}	
	\begin{align}
	p(x | \phi_{\ell},\sigma_c,\phi_{r}) & =\frac{p(\phi_{\ell},\sigma_c, \phi_{r} |x)\cdot p(x)}{p(\phi_{\ell}, \sigma_c,  \phi_{r})} \nonumber \\
	& =\frac{p(\sigma_c | \phi_{\ell}, \phi_{r}, x)\cdot p(\phi_{\ell}|\phi_{r}, x)\cdot p(\phi_{r}|x)\cdot p(x)}{p(\sigma_c | \phi_{\ell}, \phi_{r})\cdot p(\phi_{\ell}|\phi_{r})\cdot p(\phi_{r})} \label{eq:trisymm-a}	 \\
	&=\frac{p(\sigma_c | \phi_{\ell}, \phi_{r}, x)\cdot p(\phi_{\ell}|x)\cdot p(\phi_{r}|x)\cdot p(x)}{p(\sigma_c | \phi_{\ell}, \phi_{r})\cdot p(\phi_{\ell})\cdot p(\phi_{r})} \label{eq:trisymm-b}	
	\end{align}
\end{subequations}	
		\vspace{-0.7cm}

\subsection{Triphone Backward}
A different possible factorization starts with the left context, leaving the center phoneme state as the single entity~\cite{bourlard1992cdnn}.\
\begin{subequations}
\label{eq:tri-bwd}	
\begin{align}
p(x | \phi_{\ell},\sigma_c,\phi_{r})&=\frac{p(\phi_{\ell},\sigma_c, \phi_{r} |x)\cdot p(x)}{p(\phi_{\ell}, \sigma_c,  \phi_{r})} \nonumber \\
& =\frac{p(\phi_{\ell}|\sigma_c, \phi_{r}, x)\cdot p(\phi_{r}| \sigma_c, x)\cdot p(\sigma_c | x) \cdot p(x)}{p(\phi_{\ell}|\sigma_c, \phi_{r})\cdot p(\phi_{r}| \sigma_c)\cdot p(\sigma_c)} \tag{7}
\end{align}
\end{subequations}

\section{Modeling and Training Details} 
\label{sec:impl} 

\subsection{Model Architecture}

The architecture of each model is divided into two separate constituting parts: (1) a bidirectional long short-time memory~(Bi-LSTM) network which obtains an encoding of the input features following the relatively well-established acoustic modeling background  \cite{graves2013hybrid,zeyer2017comprehensive}, (2) a factorized neural network which integrates the context into the whole model.\ Regarding the CD component, there are three aspects to be underlined.\ The left and right phonemes of each triphone along with the center phoneme state are represented by using an embedding layers.\ 
Each output layer is preceded by a multi-layer perceptron~(MLP). It is possible to organize the MLP layers with different settings.\ They could be used as a shared combination component or be located in independent flows.\ Experimental results over different architectures show that in case of backward and forward model, these arbitrary choices do not implicate significant changes in the model performance.\

\subsection{Multi-Stage Phonetic Training}
\label{subsec:pre}
The final models are improved by using pre-training.\ The whole procedure can be defined as a multi-stage training which builds on incremental phonetic dependencies.\ We start with a monophone network and at each stage augment the context-dependency relations to adhere to a higher acoustic-phonetic n-gram scheme.\ 
 From one stage to the following, the outputs needed for the final factorized model are kept without further modification.\ For this set of pre-training experiments on the Switchboard task, and reported in \cref{tab:pretrain}, we took advantage of the fact that the diphone model is actually a complete sub-architecture of the triphone model with the forward decomposition.\ The experiments outcome show that with no pre-training it is possible to obtain the same performance of the baseline model using CART.\ The comparison between Experiments ($\tre$) with only diphone and ($\quattro$) with only monophone pre-training stages, against Experiments ($\cinque$) and ($\sei$) shows that the model benefits from the three-stage training.\ For the proposed diphone and triphone models the WER is consequently decreased from $14.8\%$ to $14.2\%$ and from $13.9\%$ to $13.6\%$, respectively.\ Furthermore, the optional inclusion of the additional factor $p(\phi_r | \sigma_c, x)$ during the diphone stage of Experiment ($\sei$), boosts both diphone and triphone models' performance.\ For more details about the experimental setup, see \cref{sec:exp}.\


\section{Experimental Setting and Results} \label{sec:exp}
We compare the CD acoustic models described in \cref{sec:deomp} with a baseline hybrid model using the standard state-tying with CART.\ All models are trained and evaluated over 300h Switchboard-1 Release 2 (LDC97S62)~\cite{godfrey1992switchboard} and Hub5’00 data (LDC2002S09), respectively, with the aid of RETURNN and RASR toolkits~\cite{zeyer2018returnn,wiesler2014rasr}.\ 
\subsection{Setting}
The frame-wise state alignment for training derives from a GMM/HMM system relying on CART.\ Our proposed approach makes use of a state inventory consisting of 136 state labels corresponding to 45 phonemes with three states and the single-state silence entity.\ For the baseline model, a set of 9001 CART labels are considered.

 Both baseline and proposed CD models use a Bi-LSTM encoder comprising 6 forward and backward layers of size 500 with $10\%$ dropout probability~\cite{srivastava2014dropout}.\ The input speech signal to the encoder is represented by 40-dimensional Gammatone Filterbank based features~\cite{schluter2007gammatone}, optionally concatenated with i-vectors of dimension 200 for a subset of the conducted experiments~\cite{kitzaivec}.\
 All models share the same set of training hyper-parameters and are trained with frame-wise cross-entropy criteria and Adam optimizer with Nesterov momentum~\cite{dozat2016incorporating}.\ By means of Newbob scheduling, the initial learning rate of $5\times 10^{-4}$ with a decay factor of $\sqrt{0.8}$ is controlled by using the cross-validation frame error rate and decreased to a minimum value of $5\times 10^{-6}$.\ For the regularization, an $L_2$ weight decay with constant $0.01$, gradient noise with a variance of $0.3$ and the focal loss factor of $2.0$ are adopted~\cite{neelakantan2015adding,lin2017focal}.\ Each CD model is trained for 80 hours.\ The pick performance for the baseline model is obtained after 24 epochs, requiring 8.5h less than the best proposed CD model.\ Concerning the proposed approach, the one-hot encoding of the left and/or right phonemes and the center phoneme states are projected by using linear layers of dimension 10 and 30, respectively.\ Furthermore, The prior quantities appearing in the denominator of \cref{eq:dph,eq:tri-fwd,eq:trisymm,eq:tri-bwd} are estimated by averaging over the output activations of the network using a subset of the training set.\ 
 
On the recognition side, we considered both 4-gram and LSTM language models~\cite{4gram,sundermeyer2012lstm,lstm}.\ The prior scales for each factor and the LM scale are tuned by using a grid search.\ The real time factors of two baseline and forward models are compared in \cref{tab:rtf}.\ Forwarding all possible context combinations in batch mode gives an important contribution to the optimization of our approach. However, we aim to proceed with other optimization methods as a future work.

\begin{table}[t]

	\setlength{\tabcolsep}{0.8em}\renewcommand{\arraystretch}{1.2}  
	\centering
	\caption{Different real time factor~(RTF) values for comparable average number of states per frame during time synchronous prefix tree search for the triphone models with CART-based state-tying~(Base) and forward decomposition~(Fwd).}
	
	\label{tab:rtf}
	\begin{tabular}{|c|c|c|c|c|} 
		\hline
		
	     \textbf{LM} &\textbf{Model} &\textbf{\#States} &\textbf{\#Trees} & \textbf{RTF} \\ \hline

		\multirow{2}{*}{4-gram} & \multirow{1}{*}{Base} & \multirow{1}{*}{$17345$} & \multirow{1}{*}{$113$}&$0.5$ \\ \cline{2-5}
		& Fwd& $15617$ & $75$& $12.18$\\ \hline
		\multirow{2}{*}{LSTM} & \multirow{1}{*}{Base} & \multirow{1}{*}{$58504$} & \multirow{1}{*}{$179$}&$5.59$ \\ \cline{2-5}
		& Fwd& $61753$ & $199$& $13.0$\\ \hline
		
	\end{tabular}   
	
	\vspace{-0.25cm}
	
\end{table}


\subsection{Results}

The experimental results for the CD models of the proposed approach show that different decompositions obtain similar performance.\ The triphone model with forward decomposition outperforms slightly the hybrid baseline model.\ The improvement is maintained also when a different LM or i-vector adaptation are applied.\ We believe that the performance drop in case of symmetric model derives from the simplifying assumption regarding no interdependency between the two contexts, as discussed in \cref{sec:deomp}.



\begin{table}[h]
	
	\setlength{\tabcolsep}{0.4em}\renewcommand{\arraystretch}{1.2}  
	\centering
	\caption{Comparison of WERs between the baseline system using standard state-tying with CART~(Std. tying) and the proposed CD models with forward~(Fwd), backward~(Bwd) and symmetric~(sym) decompositions, using 4-gram and LSTM LMs. A subset of experiments are carried out using i-vectors~(I-vec).}
	\vspace{-2mm}	
	\label{tab:res}
	\begin{tabular}{?c|c|c?c|c|c|c?} 
		\hline
		
		\multirow{2}{*}{\textbf{Context-}} & 			\multirow{3}{*}{\textbf{LM}}&
		\multirow{3}{*}{\textbf{I-vec}} &
		\multicolumn{4}{c?}{\textbf{[\%WER]}}\\ \cline{4-7}
		\multirow{2}{*}{\textbf{Dependency}}& 
		\multirow{2}{*}{\textbf{}}& &\multirow{1}{*}{\textbf{Std.}}&\multicolumn{3}{c?}{\multirow{2}{*}{\textbf{Proposed Approach}}}  \\ 		
		& &&\textbf{Tying}&\multicolumn{3}{c?}{} \\ \hline \hline
		
				\multirow{5}{*}{\textbf{Triphone}}& \multirow{3}{*}{{4-gram}} &\multirow{2}{*}{{no}}& \multirow{2}{*}{$13.9$} & \multirow{1}{*}{\textbf{Fwd}} & \multirow{1}{*}{\textbf{Bwd}} & \multirow{1}{*}{\textbf{Sym}} \\ \cline{5-7}
		&& & & $13.6$ & $13.8$& $14.2$\\ \cline{3-7}
		&& {yes}& $13.3$& $12.9$ & $ - $& $ - $\\ \cline{2-7}

		& \multirow{2}{*}{{LSTM}}&{no}& $12.7$& $12.6$ & $12.8$& $13.8$\\  \cline{3-7}
		& &{yes}& $11.9$& $11.7$ & $ - $& $ - $\\  \hline

		\textbf{Diphone}& \multirow{2}{*}{{4-gram}} & \multirow{2}{*}{{no}} & $15.0$ & \multicolumn{3}{c?}{$14.2$} \\ \cline{1-1}\cline{4-7}


		\textbf{Monophone} & & & \multicolumn{4}{c?}{$17.3$} \\ \hline
	\end{tabular}   
	
	
\end{table}

 \section{Discussion}
 
For the proposed CD models, the identity of each HMM state is uniquely defined by the identity of the center phoneme and the position within it, along with its right and left phonemes.\ The state labels in this case correspond to the set of CI states.\ The consideration of a subset of factors and not the full factorized model during the decoding leads to a considerable performance degradation.\ As an example, for a symmetric model, if we use only the normalized posterior $\frac{p(\sigma_c | \phi_{\ell}, \phi_{r}, x)}{p(\sigma_c | \phi_{\ell}, \phi_{r})}$ from \cref{eq:trisymm}, we observe up to $48\%$ relative WER deterioration.\ Furthermore, by including an additional target belonging to a larger context span during the training and choosing a subset of the factors during decoding it is possible to obtain improvement.\ This is for example the case of the pre-trained diphone model of Experiment $\sei$ of \cref{tab:pretrain} having also the $p(\phi_r | \sigma_c, x)$ factor during training, against the pre-trained diphone of Experiment $\cinque$.\ These observations suggest two aspects about the CD models: (1) the model learns the context-dependencies during joint training of the factors, (2) the decision rule carried out with respect to the defined theoretical framework is consistent and sound.\

\section{Conclusions}
We showed that in acoustic modeling for the hybrid approach it is possible to discard the phone clustering step.\ Our results indicate that direct modeling of context provides sufficient smoothing ability with respect to the variability in context-dependent phoneme statistics and performs as well as the former clustering-based approach.\ However, at this stage of the work, the training of the models is still based on the frame-wise alignment derived from a separate GMM/HMM system.\ Future work concentrates on training this direct modeling approach from scratch, in order to also eliminate this secondary dependence on phonetic clustering.
 
\section{Acknowledgements}

This work has received funding from the European Research Council (ERC) under the European Union's Horizon 2020 research and innovation programme (grant agreement No 694537, project "SEQCLAS") and from a Google Focused Award.\ The work reflects only the authors' views and none of the funding parties is responsible for any use that may be made of the information it contains.\ We thank Markus Kitza for providing us with the i-vectors for model training.

\bibliographystyle{IEEEtran}
\bibliography{mybib}
\end{document}